# PhET: Percepciones y contribución del uso de simulaciones en el aprendizaje de los conceptos de energía para un curso de física general de la enseñanza técnica

José E. Martínez P.

*Instituto Universitario de Tecnología del Estado Bolívar, Edf. IUTEB. Casco Histórico, Ciudad Bolívar, 8001. Venezuela. Universidad Nacional Experimental de la Fuerza Armada. UNEFA; +58 4164929985; josenriquemartinez@gmail.com*

**Resumen**

El aprendizaje de la física es considerada importante en la enseñanza técnica y para la formación del ingeniero. Sin embargo, existe evidencia empírica que los cursos de ciencias han fallado en la enseñanza de los conceptos de la física, desde la escuela hasta la Universidad. En este trabajo se explora el uso de un simulador interactivo para el aprendizaje de los conceptos de energía en un curso universitario a través de una experiencia de exploración semi-guiada. Se muestra que el proceso instructivo despierta el interés en los estudiantes y contribuye a su aprendizaje.

**Palabras Clave:** Simulador interactivo, aprendizaje exploratorio semi-guiado, aprendizaje de la física, enseñanza técnica superior

**Abstract**

The physics learning is very important to techniques teaching and the engineers' upbringing. However, there is evidence that science classes from elementary school through to university are generally failing to provide most students with an understanding of physics We have explored the use of a novel aspect of information technology: interactive simulations to learning of energy's concepts. This way, students can construct their understanding through semi-guided exploration. Research has shown this process to be an effective and engaging way to learn.

## 1. Introducción

La Física y la Matemática son las asignaturas que sustentan de forma significativa la formación del futuro técnico superior universitario (ingeniero técnico), así como la formación del ingeniero. Lo anterior se devela de una muy buena cantidad de evidencia empírica que reconocen algunos autores [1-3] y destaca la serie de "Libros Blancos", que muestran el trabajo de una Red de universidades españolas apoyados por la Agencia Nacional de Evaluación de la Calidad y Acreditación, ANECA, con el propósito de entregar los estudios y supuestos prácticos para el diseño de los títulos de grado adaptados al Espacio Europeo de Educación Superior. En ellos se muestra la coincidencia de empleadores, titulados y profesores de la importancia del conocimiento de las ciencias básicas (física, química y matemática) para la formación del ingeniero.

Esta importancia de la física está sobre la base de que esta asignatura provee de una serie de conceptos, modelos, principios y leyes necesarios para la comprensión y desarrollo de una de las competencias procedimentales importante para el diseño. Además, la física entre sus objetivos clave se encuentra el establecimiento de las relaciones entre los objetos, eventos y fenómenos del mundo real y formular las teorías y modelos interpretativos, [4]. Estos argumentos colocan el aprendizaje de la física como eje principal en la formación de ingenieros y técnicos universitarios.

No obstante, existen trabajos que relatan que infortunadamente [5], los estudiantes ingresan a la educación superior con claras deficiencias en muchos de los conceptos de ciencias que forman parte básica para la construcción de nuevas estructuras conceptuales. Además, aún y cuando se continúa con la formación en las ciencias básicas, todavía persisten fallas conceptuales y procedimentales en el área de física después de graduados. Algunos autores han dado cuenta de estos fallos, entre las que se destacan: diferencias en los modelos didácticos entre el bachillerato y la educación superior, [1]; concepciones erróneas en la estructura previa de los estudiantes al ingresar a la Universidad, [6] y la estructura conceptual compleja de la física, [7].

En este sentido, investigadores han formulado algunas teorías acerca del aprendizaje en las últimas tres décadas para dar cuenta de la naturaleza y los constructos del mismo. Estos avances en la psicología del aprendizaje revelan una naturaleza distinta a la que se tenía, destacando que el aprendizaje es situado, contextual, significativo y activo.

Ahora bien, ¿cómo desarrollar una instrucción en la enseñanza de la física que tome en cuenta la naturaleza del aprendizaje, active el reacomodo en la estructura mental del estudiante a su conocimiento previo confrontándolo a diferentes contextos, y que sea activo, dada las limitaciones reales de la presentación de los diferentes modelos que tiene la física para representar los fenómenos en ambientes áulicos?

**2. Obstáculos en la enseñanza de la física y fallos de su didáctica**

Como se destacó anteriormente, la enseñanza de la física es significativa ya que provee de los conceptos básicos, principios y leyes que contribuyen al desarrollo de las competencias actitudinales, procedimentales y conceptuales del futuro ingeniero, ingeniero técnico o técnico superior universitario. Infortunadamente existe evidencia que los cursos de ciencia, desde la primaria hasta la universidad, han fallado en proveer de las competencias básicas conceptuales y en la resolución de problemas en física, [5].

La experiencia de profesores destaca la existencia de una brecha no fácil de salvar al ingresar los estudiantes a la universidad. Los estudiantes deben adaptarse a un modelo de enseñanza y aprendizaje donde deben remediar las carencias y concepciones erróneas que traen del bachillerato. Esto trae como resultado un alto índice de fracasos y abandonos, [1].

Por otro lado, Carrascosa [6] señala que los errores conceptuales de los estudiantes que ingresan a la universidad son un obstáculo importante para el aprendizaje científico con ellos relacionados. También este autor destaca que el origen de las concepciones erróneas son: Influencia de las experiencias físicas cotidianas, influencia de la comunicación verbal, visual y escrita y los libros de texto que contienen graves errores conceptuales.

Por otro lado, Marulanda y Gómez [7] destacan la dificultad generalizada que presentan los estudiantes con el aprendizaje de la física es el resultado de la complejidad de esta ciencia relacionada con la estructura de conceptos y leyes abstractos. Esto representa un gran problema en la didáctica de la física porque existen muchos fenómenos en la naturaleza que no son observables a simple vista o reproducibles en el aula de clase; y de ser así, solamente representaría una parcialidad de la realidad.

En este sentido, siempre ha existido un reto para la didáctica de la ciencia en general y en particular para la física en representar y mostrar los fenómenos de la naturaleza y los modelos que den cuenta de su explicación a partir de mecanismos áulicos y de laboratorio.

Infortunadamente, lo anterior representa un gran obstáculo en un sistema educativo masificado y serias limitaciones para muchos países que no disponen de la tecnología y financiamiento para sostener laboratorios didácticos. Una solución a esta problemática puede buscarse en las tecnologías de información y comunicación que como se mostrará, pudieran ayudar a superar estas restricciones.

## 3. Las Tecnologías de Información y Comunicación y el uso de simuladores en la enseñanza de la física

El potencial de las Tecnologías de Información y Comunicación TIC para proveer de información, contenido y conocimiento es suficientemente admitido por organizaciones y organismos mundiales y regionales entre los que destaca: Organización de las Naciones Unidas para la Educación y la Cultura, por sus siglas en inglés, UNESCO; Banco Mundial, la Organización de Estados Iberoamericanos, OEI, la Comisión Económica para América Latina y el Caribe, CEPAL; entre otros. El pronunciamiento de estas instituciones es el resultado del reconocimiento que las TIC pueden hacer para transformar las prácticas pedagógicas dentro del aula de clase. Esta postura está fuertemente anclada en varias investigaciones en el campo empírico realizadas en los últimos años y recogidas en un informe de la UNESCO [8].

Por otro lado, Sigalés [9] reseña en términos generales que, en la modalidad presencial de la actividad docente, las TIC pueden brindar apoyo a la docencia, flexibilizando los tiempos, espacios y ritmos de trabajo. De manera particular pueden contribuir a la mejora de las representaciones del conocimiento, bien sea para aproximarse a algunas parcelas de la realidad y para simular cómo se resuelven problemas en ella o para ayudar a la comprensión de sistemas conceptuales complejos.

De manera particular, los avances en la informática educativa han colocado especial énfasis en hacer de los simuladores por computadoras herramientas conceptuales basadas en los descubrimientos en la naturaleza del aprendizaje: situado, contextual, interactivo. En la didáctica de las ciencias está cobrando particular interés el uso de las simulaciones por computadoras por el tratamiento de imágenes que proporcionan una representación dinámica del fenómeno y la interactividad que le permite al estudiante involucrarse en la variación de los parámetros de estudio.

Adams y otros [10] destacan que el uso de las simulaciones interactivas por computadoras con representaciones complejas y el uso de grafos sofisticados en el aula de clases es relativamente nuevo y la investigación en esta área para determinar su influencia en el aprendizaje del estudiante, en su interés, o cómo pueden ser rediseñados y usados para mayor efectividad es muy escasa.

En este mismo orden de ideas, Pontes [11] destaca que a pesar del interés que despiertan los simuladores por computadoras, aún siguen existiendo cuestiones relevantes en el dominio de la enseñanza de las ciencias con el uso de los simuladores con computadoras que vale la pena reflexionar y avanzar: a) análisis de las funciones educativas que pueden desempeñar los simuladores por computadoras en la enseñanza de las ciencias, b) los recursos informáticos que presentan mayor interés a los estudiantes, c) búsqueda de soluciones a los problemas educativos planteados en el campo de la didáctica de las ciencias mediante el uso de las TIC y d) el desarrollo de métodos y estrategias de trabajo docente que permitan utilizar los recursos informáticos como instrumento de aprendizaje significativo. También destaca que las simulaciones por computadoras están quedando obsoletas sin que dé respuesta y se reflexione al respecto. Este trabajo fue orientado a generar conocimiento en torno a los recursos que presentan mayor interés por parte de los estudiantes.

Ahora bien, visto la funcionalidad y características que las simulaciones por computadoras tienen para proveer de interactividad, exploración y experimentación de los fenómenos de la naturaleza, mostremos los avances en el ámbito del aprendizaje y que los simuladores pueden apoyar.

## 3. Marco teórico y conceptual del aprendizaje.

Una serie de avances en la psicología del aprendizaje ocurridas en las últimas décadas revelan una naturaleza distinta a la que sustenta el modelo transmisivo, se ha revelado que la naturaleza del aprendizaje es activo, situado, contextual, interactivo, significativo y colaborativo.

En contraste con el paradigma tradicional de enseñanza-aprendizaje, ha ido emergiendo un nuevo paradigma basado en tres décadas de investigación, que abarca los siguientes conceptos sobre el proceso de aprendizaje. En la mayoría de las actividades humanas, los individuos se enfrentan al desafío de producir conocimiento y no simplemente reproducirlo. Para permitir que los alumnos alcancen niveles óptimos de competencia, deben ser motivados a involucrarse de forma activa en el proceso de aprendizaje, en actividades que incluyan resolver problemas reales, producir trabajos escritos originales, realizar proyectos de investigación científica (en lugar de simplemente estudiar acerca de la ciencia), dialogar con otros acerca de temas importantes, realizar actividades artísticas y musicales y construir objetos. El plan de estudios tradicional requiere que los alumnos únicamente recuerden y describan lo que otros han realizado y producido. Si bien toda la producción de conocimiento debe estar basada en la comprensión de un conocimiento anterior, la mera reproducción de conocimiento, desconectada de su producción, es mayormente una actividad pasiva que no involucra de modo significativo al alumno ni le presenta ningún desafío, [12].

El proceso didáctico utilizado en las escuelas actuales está basado en la noción de que cerebro funciona como un procesador en serie, planificado únicamente para procesar una unidad de información por vez, siguiendo un orden secuencial. No obstante, en realidad no es cierto, la mente es un maravilloso procesador paralelo, que puede prestar atención y procesar muchos tipos de información simultáneamente, [13]. La teoría e investigación cognitiva ve el aprendizaje como una reorganización de las estructuras de conocimiento, en este sentido el ser humano está en capacidad de atención con amplios márgenes. Las estructuras de conocimiento se guardan en la memoria semántica como esquemas o mapas cognitivos, por lo que los estudiantes atienden a diferentes estímulos simultáneamente. En tal sentido, los alumnos "aprenden" al ampliar, combinar y reacomodar un grupo de mapas cognitivos, que muchas veces se superponen o están interconectados por medio de una compleja red de asociaciones.

En este sentido, Sánchez [14] pone el acento en el modelo de Johnson-Laird (1983, 1990) que supone que el conocimiento se representa y se asimila dependiendo de la situación. Esto abre una perspectiva diferente en los otros modelos anteriores que de forma prescriptiva intentan el reemplazo de un concepto erróneo a uno nuevo. Lo anterior coloca la instrucción en el aula de clase en procurar cambios situados y contextuales y no en los cambios conceptuales. Se trata de una redimensión en la naturaleza del cambio en el aula de clase: del conceptual al situacional. Se busca un reacomodo de la estructura cognoscitiva de estudiante como consecuencia del aprendizaje situado. La implicación didáctica de este modelo prescribe la confrontación del estudiante con una variedad de materiales y situaciones que obligarían la activación de diferentes modelos mentales para facilitar el reacomodo de la estructura cognoscitiva como consecuencia del aprendizaje situado.

En este orden de ideas, las implicaciones educativas dentro del aula de clase son variadas. Para un maestro o profesor ya les he imposible mantener la atención de sus alumnos exponiendo verbalmente el contenido de una lección. Ese contenido ya está en la Internet en variadas y diversas fuentes. Allí el estudiante puede interactuar con otros pares, exponerse a los contenidos de forma textual, auditiva y visual, y en combinación, para resultar un multiestímulo que atraiga y le reanime el interés por el aprendizaje. Se debe estimular la motivación por el aprendizaje interactivo y deje el proceso de enseñanza pasivo y por transmisión.

Es por ello que las simulaciones por computadoras cobran especial relevancia en virtud de sus características y potencialidades en apoyar la naturaleza activa de quien aprende, la interactividad con el conocimiento previo y el reacomodo de nuevas estructuras cognitivas para la incorporación de un nuevo conocimiento, la visualización del fenómeno natural y los modelos gráficos que puede representar y la variedad del contexto que puede simular.

Teniendo en cuenta la potencialidad del uso de los simuladores por computadoras y la naturaleza del aprendizaje se pasa a describir el contexto de la innovación en la práctica pedagógica donde se desarrolló la innovación educativa.

## 4. Descripción de la innovación pedagógica

De manera particular, la introducción de las TIC no ha pasado de ser simplemente para apoyar la labor académico-administrativa de los profesores. No obstante, el aumento de la oferta en abierto de objetos de aprendizaje y la presión social de pensadores y constructores de una sociedad del conocimiento están dando cuenta de la necesidad de innovar en la práctica educativa en nuestras instituciones.

El Instituto Universitario de Tecnología del Estado Bolívar fue creado en el año 2002 con el propósito de administrar los programas de formación de técnicos superiores universitarios (ingenieros técnicos) en las áreas de mecánica, sistemas industriales, electricidad y geología y minas, y se encuentra ubicado en Ciudad Bolívar, Venezuela

A pesar de muy buenas intenciones de crear una Institución modelo, poco se está haciendo por incorporar las tecnologías a las prácticas educativas. Sin embargo, un grupo de profesores se dieron a la tarea de brindar y ofrecer una demanda en el uso de las tecnologías en algunas asignaturas relacionadas con las ciencias básicas. En el caso de la física, se procedió a buscar, evaluar, seleccionar e incorporar un grupo de simuladores que brindaran el apoyo didáctico áulico, por la potencialidad que ellos representan. Durante este proceso, y dado el avance de la tecnología del *software* permitiendo la incorporación de elementos que hacen a los simuladores más cercanos a la realidad, se decidió seleccionar el proyecto *Physics Education Technology* (PhET), de la Universidad de Colorado USA. Este proyecto ha puesto en abierto una diversidad de simulaciones en física para contribuir como recurso a la enseñanza y el aprendizaje de la física para cursos iniciales de física a nivel superior, disponible en http://phet.colorado.edu/new/index.php. Una principal ventaja que prevaleció para su selección es que el simulador puede descargarse en cualquier computadora y utilizarse sin necesidad de tener una conexión de Internet.

Estos simuladores comenzaron a incorporarse al aula como apoyo didáctico en clases demostrativas por parte del docente en el curso de Física General, desde su colocación en abierto desde 2005. En el semestre II de 2007 se decidió su uso sistemático para que los estudiantes los usaran de forma instrumental. En esa experiencia se pidió, de forma voluntaria, las opiniones y percepciones en el uso instrumental del simulador *"Energy Skate Park"* como contribución del aprendizaje de los conceptos de energía potencial gravitatoria, energía cinética y energía térmica y el principio de conservación.

Este simulador tiene las características descritas en uno de los apartados anteriores. Provee de la visualización dentro de una situación real. Un patinador sobre una pista prediseñada se coloca para dejarlo deslizar sobre ella. El estudiante puede cambiar a voluntad el tipo del patinador y su masa, colocarle fricción a la pista, diseñar diferentes pistas, colocar la pista en diferentes ambientes situados: en el planeta Tierra, en la Luna, en Júpiter y en el Espacio. El simulador permite al estudiante visualizar los grafos de los diferentes tipos de energía y su relación con la posición y el tiempo en forma dinámica.

Por otra parte, el simulador tiene un diseño ambientado a la realidad y permite seleccionar las diferentes localizaciones donde se puede visualizar el entorno que se recrea (la Luna, el Espacio, Júpiter). El estudiante puede ver y contextualizar el ambiente donde se encuentra la persona o el objeto, puede recrear con herramientas de dragado y pegado, y diseñar diferentes pistas y visualizar, explorar, contextualizar y experimentar. Además, el estudiante puede colocar el sistema de coordenadas y ajustar el nivel de referencia de la Energía Potencial Gravitatoria. En la figura 1 se observa las características y ventajas del simulador, y los diferentes parámetros con que el estudiante puede experimentar.

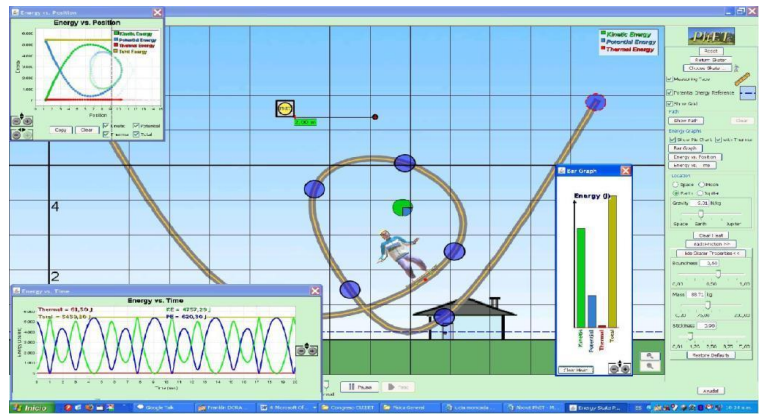

*Figura 1.* "Energy Skate Park"

## 4.1 Diseño didáctico utilizando como herramienta el simulador por computadora *"Energy Skate Park"*

El diseño didáctico para desarrollar los contenidos relacionados con los conceptos de Energía Potencial Gravitatoria, Energía Cinética, Energía Térmica y la Conservación de la Energía fue desarrollado en tres fases.

Primera fase: En una clase demostrativa áulica se mostró a los estudiantes el simulador. El docente era quien manejaba el simulador y se proyectó a la audiencia para que ellos observaran las características más relevantes. No se entregó ningún texto explicativo donde se dieran las definiciones formales de los contenidos y conceptos mencionados. El docente inició con preguntas abiertas para inducir la formulación de hipótesis donde a partir de un guión preestablecido, los estudiantes en grupos de tres consultaban entre ellos y luego se abría el debate para su comprobación. La intención de este diseño fue promover el aprendizaje colaborativo y exploratorio.

Segunda fase: En una segunda fase del diseño, se preparó un video digital instructivo donde se le indicaba al estudiante el uso instrumental del simulador. Este video se colocó en la plataforma de *Moodle* donde se tiene desarrollado todo el contenido de la asignatura.

Tercera fase: En una última fase se les entregó a los estudiantes un cuestionario de preguntas inquisitivas donde se intentó promover el aprendizaje exploratorio individual, preferiblemente. Se pretendió orientar la construcción y reacomodo del conocimiento a su estructura previa. Está diseñado para: a) cuestionar algún conocimiento previo erróneo de posea el estudiante de origen escolar y b) fomentar la interacción entre el estudiante y el simulador. En el caso del concepto de Energía Potencial Gravitatoria se orienta al estudiante para que indague cambios por modificaciones en la masa, en la altura y en la gravedad, (experimentando colocar al patinador en los distintos ambientes que posee el simulador). El estudiante construye sus conceptos y relaciones y los trae al debate en el aula de clases.

## 5. Metodología y presentación de resultados.

Sobre la base del objetivo propuesto, la matriz epistémica que sirvió de plataforma fue la concepción fenomenológica [15]. Esta selección tuvo como intención interpretar, sobre la base de un cuestionario de preguntas abiertas, las percepciones de los alumnos acerca de la contribución del simulador usado en el aula para la reconstrucción de los conceptos antes señalados.

Se pidió voluntariamente que respondieran los cuestionarios. Participaron 42, de un total de 120 del programa de formación de técnicos superiores universitarios de Mecánica cursantes de la asignatura de Física General administrada durante el periodo semestre 2007-II, representando un 35%.

El diseño del cuestionario de preguntas abiertas buscaba hacer emerger la representación de los significados que los estudiantes tuvieron del simulador. Estas preguntas permitieron a los estudiantes expresarse acerca de la visualización, contexto e interactividad. Además, se indagaron aspectos relativos con las funciones del simulador: exploración, investigación y experimentación

Por otra parte, también se construyó un cuestionario con preguntas de selección en una escala tipo Likert para indagar la contribución de la simulación a su aprendizaje. En este cuestionario se indagó aspectos relacionados con la usabilidad y contribución del simulador al aprendizaje de los conceptos relacionados con la Energía. La usabilidad, en este caso, buscaba valorar la opinión del estudiante con relación a la facilidad de uso del simulador para inducir la construcción de los conceptos de energía y su conservación y la contribución, indaga la opinión que tiene el estudiante de ese uso para construir los conceptos de energía y su conservación.

La aplicación de esta metodología arrojó los siguientes resultados en cuanto al cuestionario aplicado de preguntas cerradas. Como se muestra en la figura 2, un 89% de los estudiantes se muestran inclinados por indicar que el simulador tiene una gran facilidad de uso, ya que un 36% lo cataloga como "Mucho" y un 43% como "Totalmente".

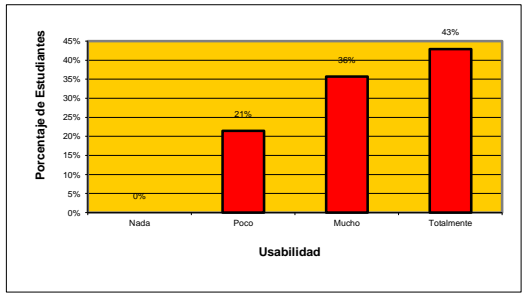
*Figura 2. Opinión de los estudiantes en cuanto a la facilidad de uso (N=42)*

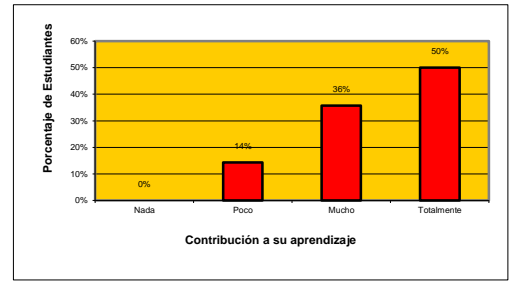
*Figura 3. Opinión de los estudiantes en relación a la contribución a su aprendizaje (N=42)*

En cuanto a la opinión de los estudiantes en relación a la contribución que tiene el simulador para la construcción de los conceptos de Energía Potencial Gravitatoria, Energía Cinética, Energía Térmica y La Conservación de la Energía; se observa en la Figura 3 que el 86% de los estudiantes considera que contribuye al aprendizaje de tales conceptos y sus relaciones con otros parámetros, distribuidos de esta manera: un 36% opina que contribuye "Mucho" y un 50% que contribuye "Totalmente".

Por otro lado, en relación al cuestionario de preguntas abiertas, se aplicó un procedimiento de tres pasos: a) Se transcribieron los contenidos de los 42 estudiantes expresados en comentarios, opiniones y respuestas a las preguntas, b) se dividieron los contenidos en unidades temáticas o párrafos que expresan una idea o concepto central y c) se establecieron categorías para darles el significado a las expresiones verbales que los participantes definen del objeto [15].

Como resultado de este proceso se puede destacar que los estudiantes expresan categorías de: favorable al aprendizaje de los conceptos de energía, ayuda a la definición cualitativa y cuantitativa de los conceptos, permite el manejo y diseño del recorrido, es como un juego, admite la visualización de los gráficos comparando las variaciones de los parámetros, recrea en diferentes ambientes con distintos valores de la aceleración de la gravedad, despierta el interés y motiva el aprendizaje.

**6. Conclusiones**

A partir de los significados de las opiniones expresadas por los estudiantes se puede concluir que el simulador por computadora *"Energy Skate Park"* contribuye favorablemente, desde la perspectiva de los estudiantes, a la reconstrucción de los conceptos de energía. Además, los estudiantes también perciben que el simulador fue de gran utilidad. Además, los estudiantes lo catalogan como de fácil uso y útil para la construcción de los conceptos de Energía.

**4. Referencias**